\documentclass{article}
\usepackage{graphicx} 
\usepackage{amsmath}
\usepackage{amsfonts}
\usepackage{hyperref}
\usepackage{booktabs}
\usepackage{bbm}
\usepackage{afterpage}
\usepackage{float}
\usepackage{placeins}

\title{Robust Inference of Dynamic Covariance Using Wishart Processes and Sequential Monte Carlo}
\author{\small Hester Huijsdens$^{1*}$, David Leeftink$^1$, Linda Geerligs$^1$, and Max Hinne$^1$ \\ \\
 \small $^1$ Donders Institute, Radboud University, Nijmegen, The Netherlands \\
 \small $^*$ Correspondence: hester.huijsdens@donders.ru.nl}
\date{} 

\newcommand{\cov}{\boldsymbol{\Sigma}} 
\newcommand{\Dof}{v}
\newcommand{\Variables}{d}
\newcommand{\Observations}{n}
\newcommand{\Particle}{s}
\newcommand{\Nrinducing}{w}

\begin{document}

\maketitle

\begin{abstract}
    Several disciplines, such as econometrics, neuroscience, and computational psychology, study the dynamic interactions between variables over time. A Bayesian nonparametric model known as the Wishart process has been shown to be effective in this situation, but its inference remains highly challenging. In this work, we introduce a Sequential Monte Carlo (SMC) sampler for the Wishart process, and show how it compares to conventional inference approaches, namely MCMC and variational inference. Using simulations we show that SMC sampling results in the most robust estimates and out-of-sample predictions of dynamic covariance. SMC especially outperforms the alternative approaches when using composite covariance functions with correlated parameters. We demonstrate the practical applicability of our proposed approach on a dataset of clinical depression ($n=1$), and show how using an accurate representation of the posterior distribution can be used to test for dynamics on covariance. \\ 

\end{abstract}

\section{Introduction}
Various domains study the joint behaviour of multiple time series. For example, in the human brain these time series consist of neuronal activation patterns, in finance they represent stock indices, and in psychology they show self-reported measures of mental health. For many research questions in these domains, it is essential to study the covariance structure between different time series. In neuroscience for example, the communication between different brain areas is studied~\cite{lurie2020questions, calhoun2014chronnectome}, which in turn can be used as a marker to diagnose several neurological disorders~\cite{Fornito2015}. Other examples include assessing the risks and returns of stock portfolios by investigating the covariance of different assets~\cite{Ledoit2003}, and investigating the co-occurrence of symptoms in mental disorders ~\cite{Borsboom2017,cramer2010comorbidity, schmittmann2013deconstructing}. Recently, there has been a shift in focus from hitherto \emph{static} representations of these interactions, to \emph{dynamic} covariances, in which the interactions between time series change as a function of an input variable. For example, recent findings in neuroscience suggest that the interactions between brain regions can change over time and that modelling the covariance between brain regions dynamically provides more sensitive biomarkers for cognition~\cite{lurie2020questions, liegeois2019resting, calhoun2014chronnectome}. Similarly, in finance the dynamic interactions between stock markets are used to study volatility and financial crises~\cite{chen2022dynamic,mollah2016equity,chiang2007dynamic,karanasos2014}. Lastly, in psychology, covariance structures between mental health markers are shown to be altered in individuals with neuroticism~\cite{bringmann2016assessing} and major depressive disorder (MDD)~\cite{pe2015emotion}. Namely, the covariance between symptoms is stronger in subjects diagnosed with major depressive disorder compared to healthy controls. Importantly, even within a single subject, the covariance structure between symptoms changes near the onset of depressive episodes, providing potential early warning signals~\cite{wichers2016critical}.

Processes of dynamic covariance can be modelled in several ways. The most prominent approaches include the multivariate generalised autoregressive conditional heteroscedastic (MGARCH) model commonly used in finance~\cite{bollerslev1986generalized, bauwens2006multivariate, brownlees2011practical, hansen2005forecast}, and the sliding-window approach that is popular in neuroscience~\cite{sakouglu2010method, allen2014tracking}. However, both approaches have a number of shortcomings. Importantly, the MGARCH-family of models requires that observations are evenly spaced over the input domain. Although even spacing is often not a problem when the questions concern time series, there are many examples where even spacing is not feasible, for example when studying cross-sectional age-related differences or how medication dosage affects the covariance of mental health symptoms. Furthermore, the sliding-window approach  requires the user to determine a number of parameters, such as window size and stride length, that can greatly affect the dynamic covariance estimates they result in~\cite{Shakil2016,Mokhtari2019}. For example, larger window sizes will result in slower observed changes in covariance, while smaller window sizes will result in more noisy measurements of covariance. In an attempt to address these challenges, Wilson and Ghahramani~\cite{wilson2010generalised} introduced the generalised Wishart process. The Wishart process is a Bayesian nonparametric approach based on Gaussian processes (GPs)~\cite{rasmussen2005} that, in contrast to the aforementioned methods, can handle unevenly spaced observations. Furthermore, as a Bayesian approach, it does not provide a point estimate of dynamic covariance, but a distribution over dynamic covariance structures. This in turn allows the model to indicate its estimation uncertainty, which enables the user to perform statistical tests. For example, with a probabilistic estimate of the Wishart process, one can test for the presence of (dynamic) covariance, even when observations are only available for a single subject. This model has been applied in different contexts, for example for modelling noise covariance in neural populations across trials~\cite{nejatbakhsh2023estimating}, when studying time-varying functional brain connectivity~\cite{kampman2024time, meng2023dynamic}, for improving the resolution in diffusion magnetic resonance imaging~\cite{cardona2015generalized}, and in combination with stochastic differential equations~\cite{jorgensen2020stochastic}.

Although the Wishart process overcomes several of the limitations of the other dynamic covariance methods, inference of the model parameters remains challenging. Especially for composite covariance functions, the model is high-dimensional, and several parameters are highly correlated. This makes the posterior distribution potentially multimodal. Wilson and Ghahramani~\cite{wilson2010generalised} inferred the model parameters using Markov Chain Monte Carlo (MCMC) sampling. Although MCMC samplers are guaranteed to converge to the true distribution, they have difficulties in sampling from high-dimensional distributions efficiently. As a solution, Heaukulani and van der Wilk~\cite{heaukulani2019scalable} proposed a variational inference approach based on sparse Gaussian processes~\cite{Bauer2016}. Although this approach is indeed much more scalable than MCMC-based approximations, it is not robust against local minima, and provides no posterior distribution over the hyperparameters of the model.

In this work, we propose a third approximate inference scheme for the Wishart process using Sequential Monte Carlo (SMC)~\cite{Chopin2020,del2006sequential}. SMC approximations were originally introduced for filtering approaches in state-space models~\cite{Kantas2009}, but more recently they have been gaining popularity as a generic approximate Bayesian inference technique~\cite{speich2021sequential,Wills2023}. Fundamentally, SMC performs a large number of short MCMC-based inference chains on different initializations of the model parameters, known as \emph{particles}, in parallel, which are then combined using importance sampling. The parallelisation makes SMC well-suited for inference of high-dimensional and multimodal distributions, as it tends not to get stuck in local optima. In addition, the computation that is required within the chains can largely be executed in parallel, which enables the algorithm to benefit from modern parallel compute hardware, such as GPUs. Here, we introduce the SMC inference scheme for the Wishart process and compare it to MCMC~\cite{wilson2010generalised} and variational inference~\cite{heaukulani2019scalable}. In most cases, SMC outperforms these approaches in terms of model fit and predictive performance. Furthermore, although variational inference tends to converge more quickly, SMC provides the full posterior at comparatively little additional running time. 

The paper is organised as follows. In Section~\ref{section:methods}, we describe the Wishart process and the three inference approaches: MCMC, variational inference, and SMC. In Section~\ref{section:experiments}, we compare the inference methods in different simulation studies that focus on capturing the true covariance process and the latent model parameters, and how accurately capturing these model parameters can be of importance of accurate out-of-sample predictions. In Section~\ref{section:empirical}, we demonstrate how the Wishart process can be used in practice by applying it to a data-set of self-reported depression symptoms~\cite{wichers2016critical}, and show how the distribution over the covariance can be used to test for dynamics in covariance. Section~\ref{section:discussion} concludes our comparison and discusses future research directions.

\section{Bayesian inference of Wishart processes}  \label{section:methods}
\subsection{Wishart processes}
\begin{figure}[tb]
    \centering
    \includegraphics[width=\textwidth]{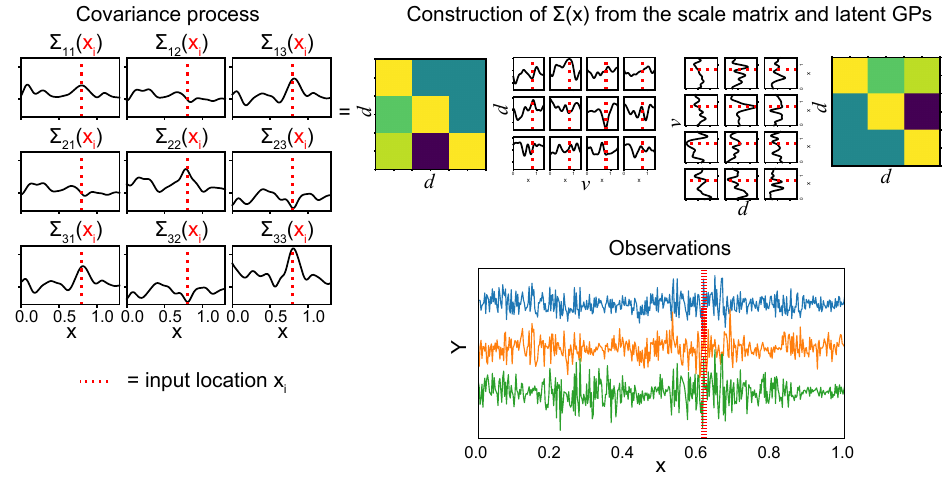}
    \caption{A visualization of the construction of the Wishart process (see Eq.~\eqref{eq:sigma_dynamic}). With $\Variables=3$ variables and $\Dof=4$ degrees of freedom, the covariance at a single input $x_i$ (represented by the red dotted line) is constructed from the outer product of the GP samples evaluated at this input. The resulting $\Variables \times \Variables$ is subsequently scaled by the $\Variables \times \Variables$ lower Cholesky decomposition of a scale matrix. Below, the time series from which to perform inference are shown.} \label{fig:construction_covariance}
\end{figure}
To understand the generalised Wishart process~\cite{bru1991wishart,wilson2010generalised, heaukulani2019scalable}, we first describe the situation in which we model a \emph{constant} covariance matrix using the Wishart \emph{distribution}. Let $\Variables$ be the number of variables (that is, time series), and let $\textbf{x} = \left( x_1, \ldots, x_\Observations \right)$ be a vector of $\Observations$ input locations, $x_i \in \mathbb{R}$, and $\textbf{Y} = \left(\textbf{y}_1, \ldots, \textbf{y}_\Observations \right)^\top$ a matrix of observations with $\textbf{y}_i \in \mathbb{R}^\Variables$, such that $\textbf{Y} \in \mathbb{R}^{\Observations \times \Variables}$. We assume that $\textbf{y}_i$ is drawn from a multivariate normal distribution with a mean of zero (although this can easily be extended to other mean vectors as well), and a covariance $\cov$:
    \begin{equation}
        \textbf{y}_i \sim \mathcal{MVN}_\Variables \left(\textbf{0}, \cov \right) \enspace, \quad i=1,\ldots, \Observations \enspace.
    \end{equation}
To learn the covariance matrix $\cov$ from the observations we follow a Bayesian approach, which implies we must decide on a prior distribution for the latent variable $\cov$.  A popular choice of prior for covariance matrices is the Wishart distribution, because it is conjugate to the normal distribution, and therefore the posterior $p\left(\cov \mid \textbf{x}, \textbf{Y} \right)$ can be computed analytically (see~\cite{zhang2021note}). The Wishart distribution is parameterised by a \emph{scale matrix} $\textbf{V}$ and a scalar \emph{degrees of freedom} parameter $\Dof$, and has the following density:
\begin{equation}
    p\left(\cov \mid \textbf{V}, \Dof\right) = \frac{|\cov|^{ \left(\Dof - \Variables - 1 \right) / 2} \exp \left(-\text{tr}\left( \textbf{V}^{-1}\cov \right) /2 \right) }{2^{\Dof \Variables/2} |V|^{\Dof / 2} \Gamma_\Variables \left(\Dof / 2 \right) } = \mathcal{W}_\Variables \left( \textbf{V}, \Dof \right) \enspace
\end{equation}
with $\text{tr}\left( \cdot \right)$ the trace function and $\Gamma_\Variables \left( \cdot \right)$ the multivariate gamma function. The intuition behind parameters $\textbf{V}$ and $\Dof$ is as follows. Suppose we have a matrix $\textbf{F} \in \mathbb{R}^{\Variables \times \Dof}$, of which each column is drawn independently from a multivariate normal distribution with a mean of zero and no covariance between the elements (that is, the covariance matrix is the identity matrix $\textbf{I}$), i.e. $\textbf{f}_l = \left( f_{1l}, \ldots , f_{\Variables l} \right)^\top \sim \mathcal{MVN}_\Variables \left( \textbf{0}, \textbf{I} \right)$, then the sum over the outer products of the $\Dof$ columns of $\textbf{F}$ is Wishart distributed with scale matrix $\textbf{I}$ and $\Dof$ degrees of freedom. Additionally, we can scale the outer products by the lower Cholesky decomposition $\textbf{L}$ of scale matrix $\textbf{V}$ (that is, $\textbf{V} = \textbf{L}\textbf{L}^\top$):
\begin{equation} \label{eq:gaussian_to_wishart}
\begin{split}
    \textbf{f}_l = \left(f_{1l}, \ldots, f_{\Variables l}\right)^\top &\sim \mathcal{MVN}_\Variables \left( \textbf{0}, \textbf{I} \right) \enspace, \quad l=1,\ldots,\Dof \\
    \cov = \sum_{l=1}^\Dof \textbf{L} \textbf{f}_l \textbf{f}_l^\top \textbf{L}^\top &\sim \mathcal{W}_\Variables \left( \textbf{V}, \Dof \right) \enspace, \qquad l=1,\ldots,\Dof \enspace.
\end{split}
\end{equation}
The resulting covariance matrix $\cov \in \mathbb{R}^{\Variables \times \Variables}$ is Wishart distributed with scale matrix $\textbf{V}$ and degrees of freedom $\Dof$.

\begin{figure}[tb]
    \centering
    \includegraphics[width=0.8\textwidth]{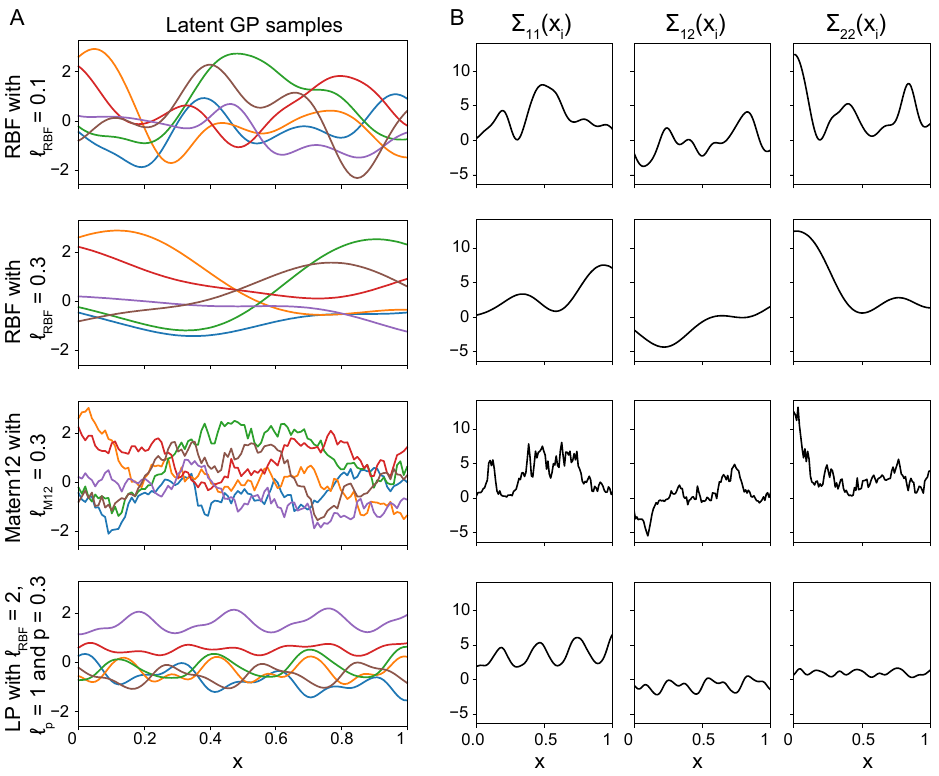}
    \caption{Different GP covariance functions can model different covariance structures. (\textbf{A}) GP samples drawn from GP priors with different covariance functions and hyperparameters. The GP samples have an RBF covariance function with $\ell_{\text{RBF}}=0.1$ or $\ell_{\text{RBF}}=0.3$, a Mat\'ern 1/2 covariance function with $\ell_{\text{M12}} = 0.3$, or a Locally Periodic covariance function with $\ell_{\text{RBF}}=2.0$, $\ell_{\text{p}}=1$ and $p=0.3$. (\textbf{B}) The covariance processes are constructed from the GP samples on the left and the lower Cholesky decomposition of the scale matrix, $\textbf{L}$ (here set to the identity matrix). The upper triangular elements of $\cov\left(x_i\right)$ are visualised, showing that the covariance process is a complex combination of GP samples.}
    \label{fig:latent_gps_and_covariance}
\end{figure}

In the Wishart \emph{process}, the constant covariance matrix $\cov$ is replaced by an input-dependent covariance matrix. This leads to the following definition of the observations, where now $\cov$ is parameterised by $x$:
\begin{equation} \label{eq:y_multivariate}
    \textbf{y}_i \sim \mathcal{MVN}_\Variables \left(\textbf{0}, \cov \left( x_i \right) \right) \enspace, \quad i=1,\ldots,\Observations \enspace.
\end{equation}
Wilson and Ghahramani~\cite{wilson2010generalised} describe a constructive approach for $\cov\left(x_i\right)$ that is similar to how the Wishart distribution is made out of normal distributions. This time the multivariate normal distributed columns of $\textbf{F}$ in Eq.~\eqref{eq:gaussian_to_wishart} are replaced by i.i.d. GPs evaluated at $\textbf{x}$. These GPs have a zero mean function and a kernel function $\kappa_\theta$, with $\theta$ as its hyperparameters:
\begin{equation}
    \textbf{f}_{jl}\left(x\right) \sim \text{GP}\left(\textbf{0}, \kappa_\theta\right) \enspace, \qquad j=1, \ldots, \Variables \enspace, \qquad l=1, \ldots, \Dof \enspace.
\end{equation}
Under the assumption that $\kappa_\theta\left(x_i, x_i\right) = 1$, taking the sum of outer products for every observation $x_i$ results in a Wishart distributed covariance matrix, that is $\cov(x_i) = \sum_{l=1}^\Dof \textbf{f}_l\left(x_i\right) \textbf{f}_l\left(x_i\right)^\top \sim W_\Variables \left(\textbf{I}, \Dof \right)$, where $\textbf{I}$ is the identity matrix. Similar to the Wishart distribution, we then construct $\cov$ by scaling this sum of outer products by the scale matrix $\textbf{V}$:
\begin{equation} \label{eq:sigma_dynamic}
    \cov\left(x_i\right) = \sum_{l=1}^\Dof \textbf{L} \textbf{f}_l \left(x_i\right) \textbf{f}_l \left(x_i\right)^\top \textbf{L}^\top \sim W_\Variables \left( \textbf{V}, \Dof \right) \enspace, \qquad i=1,\ldots,\Observations \enspace,
\end{equation}
with $\textbf{V} = \textbf{L}\textbf{L}^\top$ as before. Fig.~\ref{fig:construction_covariance} provides a visual illustration of the constructive approach to the Wishart process. 

To complete the Bayesian model, we define the following prior distribution scheme. We set a normal prior on each element of $\textbf{L}$ independently, and determine the prior of $\theta$ based on the covariance function (as will be described in Section 3):
    \begin{align*}
    \theta &\sim p(\theta) & \\
    \textbf{f}_{jl} &\sim \text{GP} \left(\textbf{0}, \kappa_{\theta} \right) \quad \text{and } \textbf{f}_l = \left( f_{1l}, \ldots, f_{\Variables l} \right)^\top &\quad j&=1, \ldots, \Variables \enspace, \quad l = 1, \ldots, \Dof \\
    L_{jo} &\sim \mathcal{N} \left( 0, 1 \right) &\quad j&=1,\ldots, \Variables  \enspace, \quad o=1,\ldots,\Variables \\
    \cov\left( x_i \right) &= \sum_{l=1}^\Dof \textbf{L} \textbf{f}_l \left( x_i \right) \textbf{f}_l \left( x_i \right)^\top \textbf{L}^\top &\quad i&=1, \ldots, \Observations \\
    \textbf{y}_i &\sim \mathcal{MVN}_\Variables \left(\textbf{0}, \cov \left( x_i \right) \right) &\quad i&=1, \ldots, \Observations \enspace.
    \end{align*}

Finally, if we want to predict observations $\textbf{y}_*$ at test locations $\textbf{x}_*$, we first predict the latent GPs: 
    \begin{equation}
        \textbf{f}_{jl}^* \mid \textbf{f}_{jl} \sim \mathcal{MVN}_{\Observations_*} \left( \textbf{K}_{*x} \textbf{K}_{xx} \textbf{f}_{jl}, \textbf{K}_{**} - \textbf{K}_{*x}\textbf{K}_{xx}^{-1}\textbf{K}_{*x}^\top \right) \quad j=1,\ldots, \Variables \enspace, \qquad l=1, \ldots, \Dof \enspace,
    \end{equation}
where $\textbf{K}_{*x}$ is formed by evaluating $\kappa_\theta$ at all combinations of test and training inputs, and $\textbf{K}_{**}$ by evaluating $\kappa_\theta$ at all pairs of test locations. Subsequently, we construct the covariance process using Eq.~\eqref{eq:sigma_dynamic} and then sample $\textbf{y}_*$ using Eq.~\eqref{eq:y_multivariate}.

An important property of the Wishart process is that the covariance function $\kappa_\theta$ can be used to express different qualitative beliefs of the dynamic covariances. For example, when using the Radial Basis Function (RBF) as covariance function, the covariance process becomes autocorrelated and smooth, as covariances corresponding to nearby input locations will be similar. Alternatively, if periodicity is expected in the covariance process, we can model this using a (locally) periodic function for $\kappa_\theta$. Fig.~\ref{fig:latent_gps_and_covariance} demonstrates a few examples of how different covariance functions and hyperparameters influence the covariance process. 

\subsection{Bayesian inference}
Although the construction of the Wishart process appears to be a straightforward extension of the Wishart distribution, inference of the corresponding posterior distribution $p \left( \cov(x_i) \mid \textbf{x}, \textbf{Y} \right)$ (Note that the dependency on $\textbf{x}$ is sometimes omitted to improve legibility when no confusion is likely to arise.) is substantially more involved. Foremost, the likelihood of the Wishart process is not conjugate to the prior, which prohibits exact inference and forces us to opt for approximate methods instead. However, this remains a challenge, as some of the model parameters are highly correlated. Previous studies have sampled from the posterior using Markov Chain Monte Carlo (MCMC) sampling~\cite{wilson2010generalised}, or approximated the posterior using a variational approach~\cite{heaukulani2019scalable,nejatbakhsh2023estimating}. Here, we introduce a third method to inference of the Wishart process, based on Sequential Monte Carlo (SMC) samplers~\cite{del2006sequential}. Before expanding on this new approach, we briefly recap the existing algorithms used to infer the posterior distributions of a Wishart process. 

\subsubsection{Markov Chain Monte Carlo for Wishart processes} \label{subsection:mcmc}
We want to infer the posterior $p\left(\cov(x_i) \mid \textbf{x}, \textbf{Y}\right)$. Since $\cov\left(x_i\right)$ follows deterministically from $f_{jl}\left(x_i\right)$, for all $j \in 1, \ldots , \Variables$ and $l \in 1, \ldots , \Dof$, this comes down to learning the posterior $p\left(\textbf{F}, \textbf{L}, \theta \mid \textbf{x}, \textbf{Y}\right)$, where $\textbf{F}$ contains all $\Variables \times \Dof$ independent GP samples. Wilson and Ghahramani~\cite{wilson2010generalised} use MCMC sampling to infer this posterior distribution. More specifically, they use Gibbs sampling \cite{geman1984stochastic} where in each MCMC iteration the parameters are updated according to their conditional distributions:
\begin{equation} \label{eq:gibbseq}
    \begin{split}
        p\left(\textbf{F} \mid \theta, \textbf{L}, \textbf{x}, \textbf{Y}\right) & \propto p\left(\textbf{Y} \mid \textbf{F}, \textbf{L}, \textbf{x}\right) p\left(\textbf{F} \mid \textbf{x}, \theta\right) \enspace \\
        p\left(\theta \mid \textbf{F}, \textbf{L}, \textbf{x}, \textbf{Y}\right) & \propto p\left(\textbf{F} \mid \textbf{x}, \theta\right) p\left(\theta\right) \enspace \\
        p\left(\textbf{L} \mid \theta, \textbf{F}, \textbf{x}, \textbf{Y}\right) & \propto p\left(\textbf{Y} \mid \textbf{F}, \textbf{L}, \textbf{x}\right) p\left(\textbf{L}\right) \enspace,
    \end{split}
\end{equation}
Here, for the parameters $\theta$ and $\textbf{L}$, we use a Random Walk Metropolis-Hastings sampler, and the GPs are sampled using elliptical slice sampling~\cite{murray2010elliptical}, since the GPs are highly correlated. Elliptical slice sampling does not require any input parameters, and is efficient because new proposals are always accepted. Moreover, the elliptical slice sampler can benefit from the Kronecker structure of $p\left(\textbf{F} \mid \textbf{x}, \theta\right)$ by sampling $\textbf{F}$ for every block independently, and then combining the resulting samples. In other words, we can independently sample $\Variables \times \Dof$ GPs, and then construct $\textbf{F}$. The computational benefit is that, instead of having to invert an $\Observations \Variables \Dof \times \Observations \Variables \Dof$ matrix, we now only have to invert $\Variables \Dof$ matrices that are $\Observations\times \Observations$, which is computationally much faster.

\subsubsection{Variational Wishart processes}
The Gibbs sampler by Wilson and Ghahramani~\cite{wilson2010generalised} was shown to be effective, but scales unfavourably to higher dimensions, both in the number of observations $\Observations$ and the number of variables $\Variables$. To address this issue, Heaukulani and van der Wilk~\cite{heaukulani2019scalable} instead propose to approximate the posterior using variational inference, a method that uses optimisation instead of sampling. Additionally, they make use of several techniques commonly found in the GP literature, such as sparse Gaussian processes~\cite{hensman2013gaussian}, to make inference more efficient.

First, a variational distribution $q\left(\textbf{F} \mid \phi\right)$ is introduced over the $\Variables \Dof$ GP samples $\textbf{F}$. The variational  parameters $\phi$  are optimised so that the distribution $q$ is close to the true posterior, with the distance between two distributions being measured by the Kullback-Leibler divergence~\cite{kullback1951kullback}, denoted $\text{KL}\left(q \| p\right)$. Ideally, the form of $q\left(\textbf{F} \mid \phi\right)$ should capture the shape of the posterior, and is often chosen to follow a multivariate normal distribution with each GP sample having mean $\textbf{m}$ and covariance matrix $\textbf{S}$, such that:
\begin{equation}
    q \left( \textbf{f}_{jl} \mid \phi \right) = \mathcal{MVN}_{\Observations} \left( \textbf{m}_{jl}, \textbf{S}_{jl} \right) \enspace, \qquad j = 1,\ldots , \Variables \enspace, \qquad l = 1,\ldots, \Dof \enspace.
\end{equation}
The variational parameters $\textbf{m}_{jl}$ and $\textbf{S}_{jl}$, as well as the Wishart process parameters $\textbf{L}$ and $\theta$, are optimised by iteratively maximising the evidence lower bound (ELBO) using gradient descent:
\begin{equation} \label{eq:elbo}
    \text{ELBO} = \sum_{i=1}^\Observations \mathbb{E}_{q \left( \textbf{F}\left(x_i\right)\right)} \left[ \log p \left( \textbf{y}_i \mid \textbf{F} \left(x_i\right)\right) \right] - \sum_{j=1}^\Variables \sum_{l=1}^\Dof \text{KL} \left( q \left( \textbf{f}_{jl} \right) \| p\left( \textbf{f}_{jl} \right) \right) \enspace,
\end{equation}
where $\textbf{F}\left( x_i \right)$ is the $\Variables \times \Dof$ matrix of GPs evaluated at $x_i$. The first part of the ELBO represents the model fit, computed as the expectation of the log-likelihood, using the variational distribution. The second part of the ELBO pushes the variational distribution closer to the prior distribution via the negative KL-divergence. Maximising the ELBO means that we try to maximise the likelihood, while also keeping the variational distribution close to the prior of $\textbf{F}$ (via the KL-divergence).

To scale to larger numbers of observations, Heaukulani and van der Wilk~\cite{heaukulani2019scalable} make use of sparse GPs~\cite{hensman2013gaussian}, where instead of learning $\textbf{F}$ for every observation, $\textbf{F}$ is learned only for a small number of inducing points $\Nrinducing$ (where $\Nrinducing \ll \Observations$). This works as follows. Let $\textbf{z} = \left(z_1, \ldots, z_\Nrinducing \right)$ be the inducing locations and $\textbf{U}$ the GP evaluations at these locations. Instead of learning a variational distribution over $\textbf{F}$, we now learn a variational distribution over $\textbf{U}$, $q\left( \textbf{u}_{jl} \right) = \mathcal{MVN}_\Nrinducing \left( \textbf{m}_{jl}, \textbf{S}_{jl} \right)$, and approximate $q \left( \textbf{F} \right)$ during prediction as follows (for every $j=1,\ldots,\Variables$ and $l = 1,\ldots, \Dof$):
\begin{equation}
    \begin{split}
    q \left( \textbf{f}_{jl} \right) &= \int p \left( \textbf{f}_{jl} \mid \textbf{u}_{jl} \right) q \left( \textbf{u}_{jl} \right) \,\text{d}\textbf{u}_{jl} \enspace \\ &= \mathcal{N}
    \left( \textbf{K}_{xz}\textbf{K}_{zz}^{-1}
    \textbf{m}_{jl}, \textbf{K}_{xx} + 
    \textbf{K}_{xz}\textbf{K}_{zz}^{-1}\left(\textbf{S}_{jl} - \textbf{K}_{zz}\right) \left(\textbf{K}_{xz}\textbf{K}_{zz}^{-1}\right)^\top \right) \enspace,
    \end{split}
\end{equation}
with $\textbf{K}_{zz} \in \mathbb{R}^{\Nrinducing \times \Nrinducing}$ containing the covariance function evaluations for every pair $\left( z, z' \right)$, and $\textbf{K}_{xz} \in \mathbb{R}^{\Observations \times \Nrinducing}$ the covariance function evaluations for every combination of training and inducing location $\left(x, z \right)$. During an optimisation step, only  $\textbf{K}_{zz}$ needs to be inverted, thereby speeding up the parameter optimisation, provided $\Nrinducing\ll \Observations$. In the ELBO in Eq.~\eqref{eq:elbo}, $p\left( \textbf{f}_{jl} \right)$ and $q\left( \textbf{f}_{jl} \right)$ are now replaced by their sparse versions $p\left( \textbf{u}_{jl} \right)$ and $q\left( \textbf{u}_{jl} \right)$ and predictions at $x_i$ are made using:
\begin{equation}
    \textbf{f}_{*jl} \mid \textbf{u}_{jl} \sim \mathcal{N}\left( \textbf{K}_{*z}\textbf{K}_{zz}^{-1}\textbf{u}_{jl}, \textbf{K}_{**} - \textbf{K}_{*z}\textbf{K}_{zz}^{-1}\textbf{K}_{*z}^\top \right) \enspace.
\end{equation}
Here $\textbf{K}_{*z}$ and $\textbf{K}_{zz}$ refer to the matrices of $\kappa_\theta$ evaluated at every combination of test and inducing locations, and every pair of inducing locations, respectively. We then draw a number of samples from $p\left( \textbf{F}_* \right)$ to construct the covariance process $\cov \left( x_* \right)$.

Additionally, Heaukulani and van der Wilk~\cite{heaukulani2019scalable} show that the parameter estimation is improved by adding an additional noise term in the construction of the covariance matrix (see Eq.~\eqref{eq:sigma_dynamic}). The noise term is a diagonal matrix $\boldsymbol{\Lambda} \in \mathbb{R}^{\Variables \times \Variables}$. $\cov \left(x_i \right)$ is now constructed as follows:
\begin{equation} \label{eq:wishart_process_noise}
    \cov \left(x_i \right) = \sum_{l=1}^\Dof \textbf{L} \textbf{f}_l \left(x_i \right) \textbf{f}_l \left(x_i \right)^\top \textbf{L}^\top + \boldsymbol{\Lambda} \enspace, \qquad i = 1,\ldots, \Observations \enspace.
\end{equation}
This regularisation term can also be optimised similar to the other parameters, and improves the approximation of the gradients required in optimisation of the ELBO.

\subsubsection{A Sequential Monte Carlo sampler for Wishart processes}
\label{smc_section}
The previously described approaches for approximate inference of the Wishart process both possess a number of drawbacks. First, due to the correlations in the model parameters and potential multimodality in the posterior, depending on the choice of covariance function $\kappa_\theta$, a standard MCMC approach is inefficient and requires a large number of samples to converge. The variational inference approach by Heaukulani and van der Wilk~\cite{heaukulani2019scalable} enables scaling applications up to larger datasets, but in practice it is prone to getting stuck in local optima. Furthermore, it does not provide a posterior distribution for the hyperparameters of the model. To overcome these limitations, we here introduce a novel Sequential Monte Carlo (SMC) sampler~\cite{speich2021sequential, del2006sequential} for posterior inference. SMC samplers are efficient at sampling from multimodal distributions, because instead of initialising the parameters at a single location, they initialise a large amount of parameter sets, called \emph{particles}, and iteratively update these particles based on their fit to the observations. Additionally, the updates for the different particles can be performed in parallel. This allows for a substantial speed increase compared to the other approaches, although of course this requires the availability of parallel computation hardware, such as GPUs.

The SMC algorithm starts from an easy-to-sample density such as the prior, and incrementally lets the particles sample from more complex densities, to eventually approach the target density. The outline of the sampler is as follows. First, $\Particle$ sets of parameters (particles) $\{ \textbf{F}^{(i)}, \theta^{(i)}, \textbf{L}^{(i)} \}_{i=1}^\Particle$ are initialised by drawing them from their prior distributions. Each particle is assigned a weight, which is initially set to $w_0^{(i)}=1/\Particle$. Next, we iteratively apply a weighting, resampling, and mutation step to adapt the particles based on their fit to the observations:
\begin{enumerate}
    \item In the weighting step, particles get assigned a weight based on how well each particle fits the data using
    \begin{equation}
        w_t^{\left( i \right)} = \frac{ p_t \left( \textbf{Y} \mid \textbf{F}_t^{ \left( i \right)}, \theta_t^{\left( i \right)}, \textbf{L}_t^{\left( i \right)}\right)}{ p_{t-1}\left( \textbf{Y} \mid \textbf{F}_t^{\left( i \right)}, \theta_t^{\left( i \right)}, \textbf{L}_t^{\left( i \right)}\right)} \enspace,
    \end{equation}
    where $t$ represents the current SMC iteration, and $i$ the particle index. Depending on the implementation, the distribution $p_t$ can change for every SMC iteration $t$. This distribution will eventually form our approximation of the posterior distribution.
    \item Next, the particles are resampled with replacement in proportion to their weights. This means that particles with small weights are discarded, and particles with large weights are duplicated. 
    \item Lastly, the particles are mutated by performing a number of Gibbs cycles (see Eq.~\eqref{eq:gibbseq}) for each particle, using the tempered distribution \\ \noindent $p_t\left( \textbf{Y} \mid \textbf{F}_t^{ \left( i \right)}, \theta_t^{\left( i \right)}, \textbf{L}_t^{\left( i \right)}\right)$ as the likelihood. This avoids the risk of all particles getting identical parameters after a few iterations.
\end{enumerate}
If we set the tempered distribution to the likelihood, we will mainly explore regions with a high likelihood, because the particles are only weighted based on their fit to the data. This risks the issue known as \emph{particle collapse}, where all particles consist of the same high-likelihood values. To overcome this, we use an adaptive-tempering variant of SMC ~\cite{jasra2011inference, speich2021sequential}. Here, the distribution at SMC iteration $t$ is tempered according to
    \begin{equation} \label{eq:tempered_likelihood}
        p_t \left( \textbf{Y} \mid \textbf{F}, \theta, \textbf{L} \right) = p\left( \textbf{Y} \mid \textbf{F}, \theta, \textbf{L} \right)^{\beta_t} \enspace,
    \end{equation}
where $\beta_t$ is the temperature that dampens the influence of the likelihood. $\beta_t$ is initially set to 0. This means we initially simply sample from the prior, and $\beta$ is then gradually increased until it reaches a value of 1, at which point we sample from the posterior. The increase in temperature between two successive SMC iterations, $\Delta\beta_t$, is determined via the effective sample size of the weights, $\Particle_{\text{eff}}$. The effective sample size is a measure of particle diversity. The new temperature at every iteration is then determined by finding $\beta_t$ such that $\Particle_{\text{eff}} = a\Particle$, where $a$ is the fraction of particles that we want to be independent~\cite{Agapiou2017,Chopin2020,Herbst2014}. 

\section{Simulation studies} \label{section:experiments}
We compare MCMC, variational inference and SMC on two distinct simulations representing different scenarios of dynamic covariances. First, we describe the data generation procedure for these simulation studies and how we will evaluate each approach. Moreover, we apply the Wishart process to an empirical dataset containing momentary states of a single subject diagnosed with Major Depressive Disorder and compare these covariance process estimates to those made by a DCC-GARCH model.

\subsection{Implementation details} \label{section:implementation_details}
Gibbs MCMC sampling is implemented using the Blackjax Python library~\cite{lao2022blackjax} which builds on top of the JAX framework~\cite{jax2018github}. As briefly mentioned in Section~\ref{subsection:mcmc}, when sampling with MCMC, we sample the covariance function hyperparameters $\theta$ and the lower Cholesky decomposition of the scale matrix $\textbf{L}$ using a Random Walk Metropolis Hastings sampler with a step size of 0.01. We use a thinning of 1000 samples and the number of burn-in steps is determined by the convergence of all model parameters, unless mentioned otherwise in the experiment. Convergence is measured by the Potential Scale Reduction Factor (PSRF)~\cite{gelman1992inference}, where we interpret a value of less than 1.1 as being converged. We measure the PSRF over four re-runs, known as `chains', of the inference algorithm, each time with a different random initialisation of the model parameters. After convergence across chains, we combine these four chains by randomly taking 250 samples of each chain.

Variational inference is implemented using the GPflow 2 library~\cite{GPflow2017}. The noise parameter in Eq.~\eqref{eq:wishart_process_noise} was initialised as $\lambda_{jj} = 0.001$ for $j=1,\ldots,\Variables$. Similar to Heaukulani and van der Wilk~\cite{heaukulani2019scalable}, we approximate the gradients of expectation of the log-likelihood in Eq.~\eqref{eq:elbo} using a small number of Monte Carlo estimates. In our results, we have used $3$ Monte Carlo estimates. To optimize Eq.~\eqref{eq:elbo}, we use the Adam optimiser~\cite{Kingma2015} with an initial learning rate of 0.001. For variational inference, we can use the PSRF only as a measure for convergence of the covariance process, but not for the latent model parameters, because variational inference only provides point-estimates for these. Therefore, we optimize the ELBO until it has not improved for 10\,000 iterations, after which we use the PSRF to check for the convergence of the covariance process. As before, the PSRF is computed over the estimates of four re-runs with a different random initialisation. Moreover, within each re-run, we inspect whether or not the latent model parameters have converged when the ELBO has converged. We then use the run with the highest ELBO for subsequent analyses. 

Similar to the MCMC implementation, SMC sampling is also implemented in the Blackjax Python library. Within each SMC cycle we sample the covariance function hyperparameters $\theta$ and the lower Cholesky decomposition of the scale matrix $\textbf{L}$ using the Gibbs MCMC sampling approach described above. We set the number of particles to 1\,000, and, unless mentioned otherwise, we base the number of mutation steps on the convergence of all model parameters, as determined by a PSRF below 1.1. This convergence is again measured over four re-runs of the inference algorithm, each time with a different random parameter initialisation. Code for our analyses is available on \href{https://github.com/Hesterhuijsdens/GWP-SMC}{GitHub}.

\subsection{Synthetic data} \label{section:data_generation}
In order to compare MCMC, variational inference, and SMC, we evaluate each approach on data with a known ground truth covariance process. In our first simulation study, we construct a dynamic covariance process that is drawn from a Wishart process prior with a Radial Basis Function (RBF; also known as the squared-exponential) as covariance function:
    \begin{equation}
        \kappa_{\text{RBF}}\left(x, x' \right) = \exp \left( - \frac{\left( x - x' \right)^2}{2 \ell_{\text{RBF}}^2} \right) \enspace. 
    \end{equation}
We set the lengthscale parameter to $\ell_{\text{RBF}} = 0.35$, representing slow dynamics, and the scale matrix $\textbf{V}$ to the identity matrix. Using $\Variables = 3$ variables and $\Dof = 4$ degrees of freedom, we draw $\Variables \times \Dof$ GP samples and construct the covariance process using Eq.~\eqref{eq:sigma_dynamic}. We repeat this covariance process generation procedure ten times while keeping the lengthscale parameter $\ell_{\text{RBF}}$ and the scale matrix $\textbf{V}$ the same. Finally, the ten resulting covariance processes are used to generate ten datasets, by sampling $\Observations = 300$ observations from a multivariate normal distribution with a mean of zero, and the latent covariance process as covariance. An example of a generated ground truth covariance process is shown in Fig.~\ref{fig:simulation_study1_estimates}A. 

In our second simulation study, we generate a covariance process that follows a rapid state-switching pattern between $\Variables = 3$ variables. This covariance process is generated as follows. The off-diagonal elements of the true latent covariance process alternate every $50$ observations between values of $0$ and $0.8$, and the variances are set to $1$. This structure is shown in Fig.~\ref{fig:simulation_study2_estimates}. Again, we generate ten datasets, but this time we use the same ground truth covariance process for all ten datasets. We draw $\Observations = 600$ observations from a multivariate normal distribution with a mean of zero and the state-switching latent covariance process, of which the first $\Observations_{\text{train}}=300$ observations are being used for inference, and the remaining  $\Observations_{\text{test}}=300$ observations for out-of-sample prediction. To capture the periodic structure of this covariance process,  we use both a Periodic covariance function and a Locally Periodic (LP) covariance function. The Periodic covariance function is defined as:
    \begin{equation} \label{eq:periodic}
        \kappa_{\text{Periodic}}\left(x, x' \right) = \exp \left( -\frac{2 \sin^2 \left( \pi |x - x'| / \textit{p} \right) }{{\ell_p}^2 } \right) \enspace,
    \end{equation}
where $p$ determines the period of the covariance, and $\ell_p$ determines the fluctuations within each period. The LP covariance function is constructed by multiplying this Periodic function with an RBF covariance function to obtain:
    \begin{equation} \label{eq:lp}
        \kappa_{\text{LP}}\left(x, x' \right) = \exp \left( -\frac{2 \sin^2 \left( \pi |x - x'| / \textit{p} \right) }{{\ell_p}^2 } \right) \exp \left(- \frac{\left( x - x' \right)^2}{2 {\ell_{\text{RBF}}}^2} \right) \enspace,
    \end{equation}
where $p$ again determines the period of the covariance, $\ell_p$ determines the fluctuations within one period, and $\ell_{\text{RBF}}$ allows the repeating covariance to change over time. Both functions should be able to capture the ground truth state switches, however the LP covariance function allows for more flexibility. We set a log-normal prior on all three parameters. 

\subsection{Performance metrics} \label{section:performance_metrics}
We evaluate the different inference approaches based on how well they recover the ground truth. Therefore, we compute the mean squared error (MSE) between the ground truth covariance process and the corresponding mean estimate of the covariance process. This metric is averaged over all $\Variables$ variables and $\Observations$ observations. Additionally, when the model parameters that were used to construct the ground truth covariance process are known (as in our first simulation), we compute the MSE between those parameters and the corresponding mean estimates. Furthermore, since the inference methods provide a distribution over the covariance, we evaluate this full distribution by computing the MSE between samples of the covariance process and the ground truth. This MSE is again averaged over all $\Variables$ variables and $\Observations$ observations. We refer to this metric over the full covariance distribution as $\text{MSE}_\text{samples}$. Hence, we use the MSE to evaluate both the accuracy of the mean covariance process estimate, and the accuracy of the distribution over the covariance process estimate.

Finally, in the second simulation study we are interested in making out-of-sample predictions. Here, we evaluate the predictive performances of the three inference methods by means of the log-likelihood (LL) of the observations given the mean covariance process estimate. This log-likelihood is averaged over the number of observations $\Observations$. To also validate the predictive performance over the entire predictive posterior distribution, we measure the Kullback-Leibler divergence (KL) between the predictive posterior distribution and the true multivariate normal distribution of the observations.

\subsection{Simulation study 1: learning the model parameters} \label{section:simulation1}
\begin{figure}[tb]
    \centering
    \includegraphics[width=\textwidth]{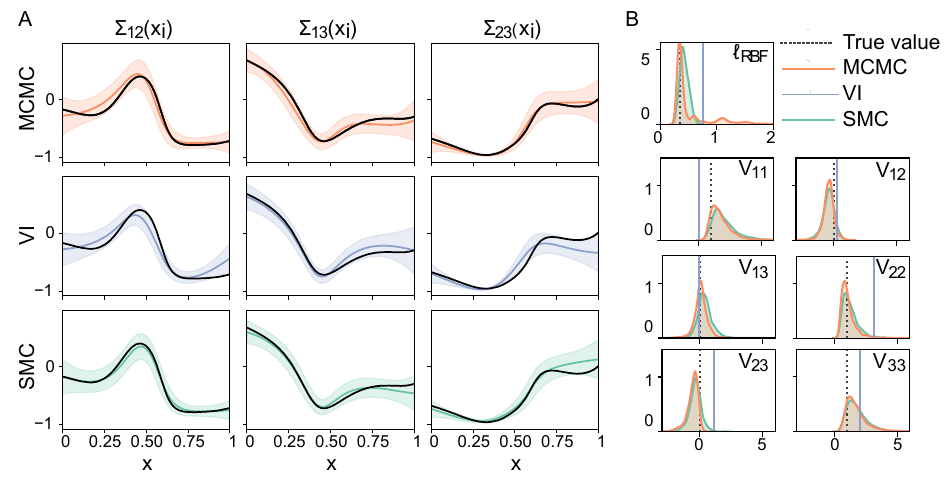}
    \caption{(\textbf{A}) Estimates of a covariance process drawn from a Wishart process prior. The ground truth covariance process is shown in black. (\textbf{B}) Inference results of the lengthscale parameter of the Radial Basis Function covariance function and the scale matrix. The true values ($\ell_{\text{RBF}} = 0.35$ and $\textbf{V} = \textbf{I}$) are indicated by the gray dotted lines.}
    \label{fig:simulation_study1_estimates}
\end{figure}
\setlength\tabcolsep{0pt}
\begin{table}[tb]
    \footnotesize
    \centering
    \caption{The accuracy of each inference method in capturing the ground truth covariance process  ($\text{MSE}_{\cov}$), lengthscale ($\text{MSE}_{\ell_{\text{RBF}}}$), and scale matrix {$\text{MSE}_\textbf{V}$} from Simulation study 1. We evaluate the mean covariance process estimate, as well as its full posterior distribution ($\text{MSE}_\text{samples}$). The computation time is shown in minutes (for MCMC per chain and for VI per initialisation). The mean and standard deviation over ten datasets are shown.\\}
    \begin{tabular*}{\linewidth}{@{\extracolsep{\fill}} lllllllllr } \toprule
         \multicolumn{1}{c}{Method} & \multicolumn{1}{c}{$\text{MSE}_{\cov}$} & \multicolumn{1}{c}{$\text{MSE}_{\ell_{\text{RBF}}}$}  & \multicolumn{1}{c}{$\text{MSE}_\textbf{V}$} & \multicolumn{1}{c}{$\text{MSE}_\text{samples}$} & \multicolumn{1}{c}{Runtime} \\ \midrule
         MCMC & 0.45 $\pm$ 0.36 & 0.00 $\pm$ 0.00 & \textbf{0.28 $\pm$ 0.26} & 0.59 $\pm$ 0.51 & 308.29 $\pm$ 7.13 \\
         VI & \textbf{0.44 $\pm$ 0.29} & 0.10 $\pm$ 0.05 & 0.55 $\pm$ 0.36 &  \textbf{0.51 $\pm$ 0.37} & \textbf{54.84 $\pm$ 0.82} \\
         SMC & 0.45 $\pm$ 0.35 & \textbf{0.00 $\pm$ 0.00} & 0.31 $\pm$ 0.28 &  0.61 $\pm$ 0.52 & 105.40 $\pm$ 2.08
         \\ \bottomrule
    \end{tabular*}
    \label{tab:model_fit_gwpprior}
\end{table}

The aim of our first simulation is to validate the ability of each inference method to capture the ground truth. Therefore, we simulate observations with a covariance process drawn from a Wishart prior (as described in Section~\ref{section:data_generation}), and use these data to measure the accuracy of MCMC sampling, variational inference, and SMC sampling in inferring the covariance process, the scale matrix, and covariance function hyperparameters. For each approach, we use the RBF covariance function, with a log-normal prior distribution on the RBF lengthscale parameter. For the current simulation study, MCMC converges after a burn-in of 4 million samples. SMC requires on average 65 SMC adaptation cycles with 2000 mutation steps within each cycle, and VI an average of 21\,420 iterations. Notably, the elements of $\cov(x)$ converge relatively quickly, while the other model parameters such as the RBF lengthscale (see Appendix~\ref{fig:convergence}) much longer to reach convergence. 

In Fig.~\ref{fig:simulation_study1_estimates}A we show the estimated dynamic covariance for one exemplar simulation run. The corresponding parameter estimates (here, the RBF lengthscale and the elements of the scale matrix $\textbf{V}$) are shown in Fig.~\ref{fig:simulation_study1_estimates}B, together with the ground truth values. MCMC and SMC both infer a distribution over the model parameters, which are visualised using kernel density estimation~\cite{scott2015multivariate}. Variational inference learns point-estimates of the model parameters, shown as a vertical line. These results are quantified using the performance measures described in Section~\ref{section:performance_metrics}, and shown in Table~\ref{tab:model_fit_gwpprior}.  

The performance measures, along with the runtime of each inference method in Table~\ref{tab:model_fit_gwpprior} indicate that all three approaches are successful in recovering the mean of the ground truth covariance process. When we look at the accuracy of estimating the latent model parameters, we see that these are recovered considerably less well by VI. In particular, MCMC and SMC both outperform VI when we look at the accuracy of inferring the RBF lengthscale and scale matrix. In other words, while variational inference reaches convergence most quickly, this comes at the cost of accurately estimating the latent model parameters, even though all three methods have converged and we took the VI result with the highest ELBO out of four re-runs (see Section~\ref{section:implementation_details}). We investigate the effect of the difference in model parameter estimation in the next simulation study.

\subsection{Simulation study 2: state switching and out-of-sample prediction} \label{section:simulation2}
\begin{figure}[!h]
    \centering
    \includegraphics[width=\textwidth]{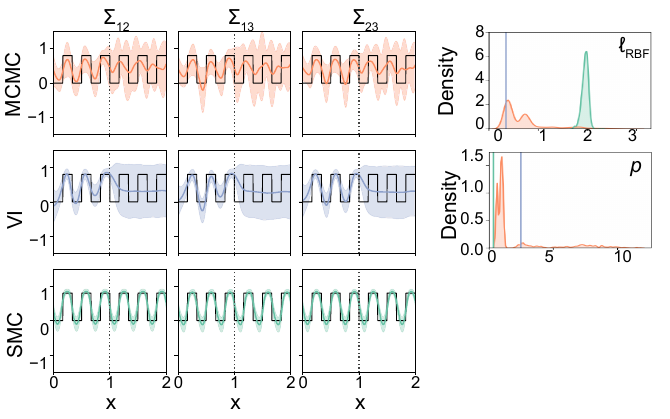}
    \caption{Covariance process estimates and out-of-sample predictions for each inference method, using a Periodic covariance function, together with the corresponding distributions over the covariance function parameters. The true covariance process is shown in black, and out-of-sample predictions are shown after the gray dotted line. The Periodic covariance function has two parameters: the period ($p$) and the lengthscale within each period ($\ell_{\text{RBF}}$).}
    \label{fig:simulation_study2_estimates}
\end{figure}

\setlength\tabcolsep{0pt}
\begin{table}[tb]
    \tiny
    \centering
    \caption{Using either a periodic or locally periodic covariance function, we show the accuracy of each inference method in capturing the mean ground truth covariance process and its full distribution. Moreover, for the out-of-sample predictions, we present the average fit to the observations and the fit of the full predictive posterior distribution. Finally, the computation time (for MCMC per chain and for VI per initialisation) is shown in minutes. The mean and standard deviation over ten datasets are shown.\\}
    \begin{tabular*}{\textwidth}{@{\extracolsep{\fill}} llllllllr } \toprule
        & \multicolumn{2}{c}{Training} & \multicolumn{4}{c}{Out-of-sample prediction} \\ \cmidrule(lr){2-3} \cmidrule(lr){4-7}
        \multicolumn{1}{l}{Method} & \multicolumn{1}{c}{$\text{MSE}_{\cov}$} & \multicolumn{1}{c}{$\text{MSE}_{\text{samples}}$} & \multicolumn{1}{c}{$\text{MSE}_{\cov}$} & \multicolumn{1}{c}{$\text{MSE}_{\text{samples}}$} & \multicolumn{1}{c}{LL} & \multicolumn{1}{c}{KL} & \multicolumn{1}{c}{Runtime}\\ \midrule
        \textit{Periodic function} & \\
         MCMC & 0.08 $\pm$ 0.02 & 0.13 $\pm$ 0.01 & 0.19 $\pm$ 0.02 & 0.34 $\pm$ 0.07 & -4.10 $\pm$ 0.07 & 0.52 $\pm$ 0.07 & 280.88 $\pm$ 4.13  \\ 
         VI & 0.06 $\pm$ 0.02 & 0.08 $\pm$ 0.02 & 0.14 $\pm$ 0.09 & 0.22 $\pm$ 0.15 & -4.01 $\pm$ 0.16 & \textbf{0.47 $\pm$ 0.20} & \textbf{30.72 $\pm$ 7.19} \\
         SMC & \textbf{0.04 $\pm$ 0.01} & \textbf{0.06 $\pm$ 0.01} & \textbf{0.11 $\pm$ 0.08} & \textbf{0.13 $\pm$ 0.09} & \textbf{-3.94 $\pm$ 0.21} & 0.48 $\pm$ 0.40 & 152.49 $\pm$ 2.11 \\ \midrule 
         \textit{LP function} & \\
         MCMC & 0.05 $\pm$ 0.01 & 0.08 $\pm$ 0.01 & 0.12 $\pm$ 0.04 & 0.24 $\pm$ 0.08 & -3.98 $\pm$ 0.13 & 0.42 $\pm$ 0.10 & 291.20 $\pm$ 3.97 \\
         VI & 0.05 $\pm$ 0.01 & 0.08 $\pm$ 0.02 & 0.10 $\pm$ 0.04 & \textbf{0.17 $\pm$ 0.08} & -3.97 $\pm$ 0.15 & 0.43 $\pm$ 0.14 & \textbf{31.62 $\pm$ 6.43} \\
         SMC & \textbf{0.04 $\pm$ 0.01} & \textbf{0.06 $\pm$ 0.01} & \textbf{0.09 $\pm$ 0.02} & 0.19 $\pm$ 0.04 & \textbf{-3.93 $\pm$ 0.08} & \textbf{0.36 $\pm$ 0.06} & 154.98 $\pm$ 2.97 \\ \bottomrule
    \end{tabular*}
    \label{tab:model_fit_states}
\end{table}
In the previous experiment, we found that all three inference methods are able to accurately estimate the covariance process, but that, unlike MCMC and SMC, variational inference did not recover the latent model parameters well. To explore how this affects the ability to make out-of-sample predictions, we use the covariance process that follows a state-switching pattern (as described in Section~\ref{section:data_generation}) to make out-of-sample predictions, and use both a Periodic and a Locally Periodic (LP) covariance function to model this covariance process. Both functions should be able to capture the state switches that are present in the ground truth, however the LP covariance function allows for more flexibility. We set a log-normal prior on all three parameters. In contrast to the RBF covariance function that was used before, which only has a single parameter, these covariance functions have two and three parameters respectively, and each of these have an important impact on out-of-sample extrapolation. For example, if the period $p$ is estimated poorly, the further away from training data we are, the more out of phase our predictions will be. Therefore, correct inference of these parameters is crucial, but this is made even more difficult due to multimodality in the posterior distribution. 

Recall that we use the first $\Observations_{\text{train}}=300$ observations for inference, and the remaining  $\Observations_{\text{test}}=300$ observations for out-of-sample prediction. Fig.~\ref{fig:simulation_study2_estimates} shows an example of an estimate and out-of-sample prediction of the three inference methods using a Periodic covariance function. Upon visual inspection, we observe that all three methods were able to estimate the periodic ground truth covariance process for the training data, although there are differences in performance. From the MSEs over the mean covariance estimate ($\text{MSE}_{\cov}$) in Table~\ref{tab:model_fit_states}, we can see that the estimates using SMC sampling outperform those made by variational inference and MCMC sampling. When looking at the MSE computed over the individual samples ($\text{MSE}_{\text{samples}}$), we observe the same pattern. On the training data, SMC sampling is most accurate in estimating the mean covariance process, and the distribution over this covariance process. Moreover, for MCMC and variational inference, we learn that the estimates using an LP covariance function were slightly more accurate than those using a Periodic covariance function.

Although the differences in performance between MCMC, variational inference, and SMC on the training data are small, they become more pronounced when looking at the predictive performance. By looking at the out-of-sample predictions in Fig.~\ref{fig:simulation_study2_estimates}, which are shown after the vertical dotted line, we observe that variational inference did not captures the periodicity in the ground truth for this set of observations. A few more estimates can be seen in Appendix~\ref{appendix:estimates_study2}. Over all sets of observations, variational inference captures the periodicity in 5/10 datasets using a Periodic covariance function, and 9/10 datasets using an LP covariance function. This is supported by the model parameter estimates in Fig.~\ref{fig:simulation_study2_estimates}, where it can be seen that the periodicity of approximately $0.33$ is not correctly inferred by variational inference. In this case, only the lengthscale parameter has learned the structure of the training data. When we look at the results of Gibbs MCMC, we see that this method too has difficulties in capturing the latent periodicity in the covariance process, especially when using the Periodic covariance function. SMC sampling gives more consistent results. Moreover, the results in Table~\ref{tab:model_fit_states} support the differences we observe in Fig.~\ref{fig:simulation_study2_estimates}, namely that the out-of-sample predictions by SMC more accurately fit the true covariance than MCMC and variational inference. Although the differences in performance between SMC and variational inference seem small when using the LP covariance function, it should be noted that we trained the model four times using different initializations for variational inference, and selected the estimates with the best ELBO.

Next, we compare the fit of the covariance estimates to the actual observations. We compare this using two metrics: the fit of the mean covariance process estimates is evaluated using the log-likelihood (LL), and the predictive posterior distribution is evaluated using the KL-divergence (KL) between this distribution and the actual distribution over the observations (see Section~\ref{section:performance_metrics}). Overall, the LL and KL-divergence results in Table~\ref{tab:model_fit_states} reveal that the differences when looking at the fit to the observations are less pronounced, although the SMC algorithm results in the best fit to the test data overall. 

Finally, we can compare the three inference methods in terms of computational efficiency. When using the Periodic covariance function, the covariance samples of MCMC have converged after a burn-in of 5\,000\,000 burn-in steps per chain, after which we collect the next 1\,000\,000 samples. We use a thinning of 1\,000 samples. However, with four chains, this means that we need to perform 24 million Gibbs cycles for all chains in total. This amount of Gibbs cycles remains the same when using the LP covariance function. It should be noted however, that the results show that MCMC still has not captured the periodic structure accurately after 5\,000 burn-in steps. SMC does capture this periodicity more accurately, and required, for both covariance functions, an average of 59 SMC adaptation cycles with 3000 mutation steps per cycle to converge. For 1\,000 particles, this means that we do 177 million Gibbs cycles in total. However, unlike for MCMC, the computations of the SMC mutation steps are parallelized across the particles, therefore greatly speeding up the algorithm. This means that only 177\,000 steps are being performed sequentially, which is much less than the 6 million steps of MCMC. Finally, VI converges much faster. Namely, VI requires an average of 46\,580 iterations per re-run until convergence of the ELBO, when using the Periodic covariance function. After convergence of the ELBO, the parameters no longer changed. Using 4 re-runs, this means we optimize the ELBO approximately 186\,320 times. When using the LP covariance function, we require 44\,700 iterations per re-run and 178\,800 iterations in total. However, although variational inference runs faster, we have seen from the results in Table~\ref{tab:model_fit_states} that variational inference has difficulty in capturing the periodicity, and therefore in making out-of-sample predictions. 

In short, these results suggest that SMC sampling can reliably capture the periodic stucture present in the data by handling the highly correlated covariance function parameters, and therefore SMC sampling can make more accurate out-of-sample predictions. From the results of Gibbs MCMC and variational inference, we can see that both methods have more difficulty with converging, and this has important consequences for out-of-sample predictions.

\section{Empirical application: dynamic correlations in depression} \label{section:empirical}
In this section we demonstrate how the generalized Wishart process can be used to study the dynamics of psychological processes, and how it enables novel analyses. Recently, there has been a paradigm shift in the study of mental disorders. Instead of defining mental disorders according to the sum score of a a set of measurements, they are now increasingly conceptualised as (dynamic) networks of interacting symptoms~\cite{bringmann2016assessing, pe2015emotion, schmittmann2013deconstructing, cramer2010comorbidity}. For example, recent studies on the onset of depressive episodes in people with Major Depressive Disorder (MDD) have shown that changes in the dynamics between individual symptoms serve as early warning signs of various mental disorders~\cite{cabrieto2019objective, wichers2016critical}. These studies modelled these dynamic correlations using a multilevel vector autoregression method~\cite{bringmann2013network}. However, we propose to apply the Wishart process in this context, because this provides us with a distribution over the covariance, which we can use to test for dynamics. Furthermore this approach allows us to work with unevenly spaced data, therefore allowing us to estimate covariance not only as a function of time, but also as a function of some other input variable, such as medication dosage.  

\subsection{Dataset, preprocessing and model choices} \label{sec:empirical_setup}
\begin{figure}[tb]
    \centering
    \includegraphics[width=0.7\textwidth]{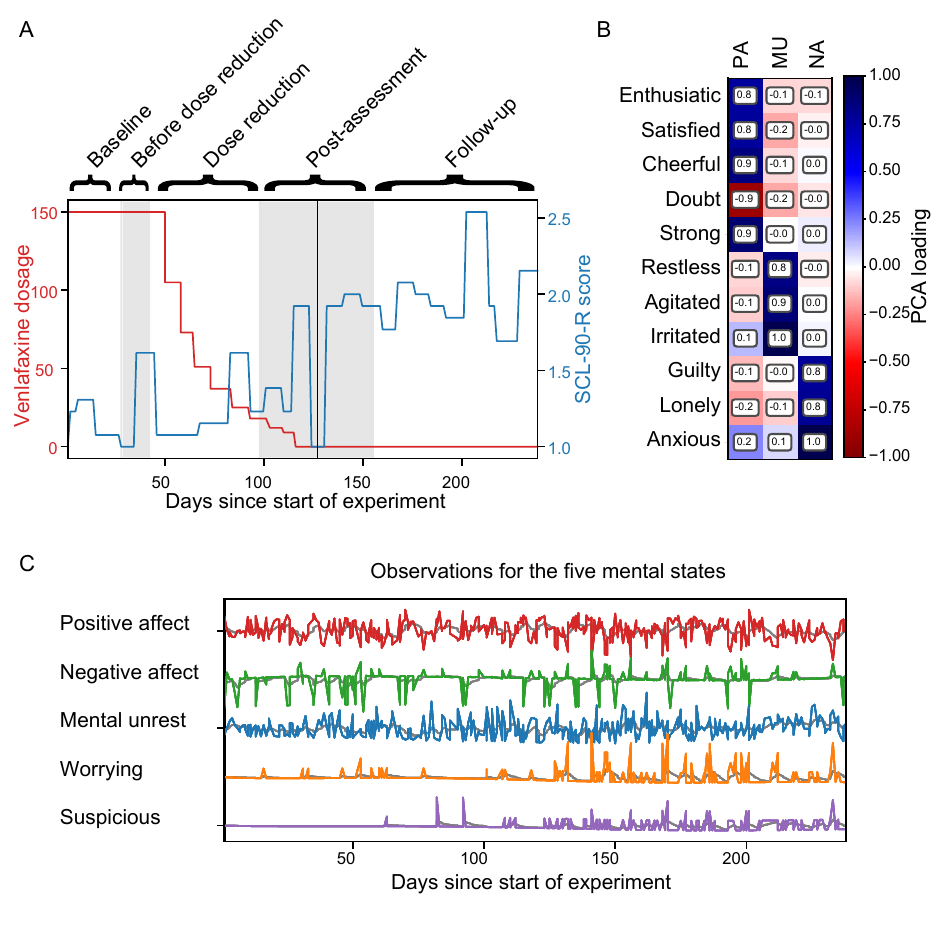}
    \caption{\textbf{(A)} Venlafaxine dosage and depression measurements, as measured by the SCL-90-R score over the course of the experiment. The five experimental phases (4 weeks baseline, 0-6 weeks before dose reduction, 8 weeks dose reduction, 8 weeks post-assessment, and 12 weeks follow-up) are shown by the shaded background, and the black vertical line indicates the moment at which the subject relapsed into a depressive episode. \textbf{(B)} The loadings from individual items to the three principal components positive affect (PA), negative affect (NA), and mental unrest (MU). \textbf{(C)} The time series data for the five mental states, together with their moving average (in gray).}
    \label{fig:empirical_mdd_data}
\end{figure}
We use the dataset from Kossakowski et al.~\cite{kossakowski2017data}, which is obtained from a single subject who has been diagnosed with MDD. The subject is a 57 year old male who monitored his mental state over the course of 237 days by filling in a questionnaire of daily life experiences several times a day. Moreover, the subject had been using venlafaxine, an anti-depressant, for 8.5 years. Interestingly, during the data collection, the dosage of venlafaxine is gradually reduced to zero in a double-blind manner according to five experimental phases: baseline (four weeks), before dosage reduction (between zero and six weeks, the exact timings unknown to the subject), during dose reduction (eight weeks), post-assessment (eight weeks), and a follow-up (twelve weeks). These phases are shown in Fig.~\ref{fig:empirical_mdd_data}A, where we can also see that the subject got more depressed over the course of the experiment, as measured on a weekly basis by the depression subscale of the Symptom Checklist Revised (SCL-90-R)~\cite{derogatis1976scl}.

Following Wichers et al.~\cite{wichers2016critical}, we collect the following items from the questionnaires: `irritated', `content', `lonely', `anxious', `enthusiastic', `cheerful', `guilty', `indecisive', `strong', `restless', and `agitated'. Subsequently, these symptoms are summarised using principal component analysis together with an oblique rotation~\cite{wichers2016critical}. The loadings of each item on these components can be seen in Fig.~\ref{fig:empirical_mdd_data}B. The components are interpreted as `positive affect', `negative affect', and `mental unrest'. Moreover, we use the items `worrying' and `suspicious' as separate variables, following again~\cite{wichers2016critical}. Finally, slow non-periodic time trends are removed from the data and, to speed up inference, only every fourth observation is kept. This results in a total of $\Observations=369$ and $\Variables=5$ variables (`positive affect', `negative affect', `mental unrest', `worrying' and `suspicious'), as visualized in Fig.~\ref{fig:empirical_mdd_data}C. 

In order to model both slow and fast changes in covariance between the five mental states, we sum an RBF and a Mat\'ern 1/2 covariance function:
\begin{equation}
    \kappa_{\text{RBF + M12}} \left(x, x' \right) = \exp \left( - \frac{\left( x - x' \right)^2}{2 \ell_{\text{RBF}}^2} \right) +  \exp \left( -\frac{| x - x'|}{2\ell^2_{\text{M12}}}  \right) \enspace.
\end{equation}
Moreover, to model the slow fluctuations of the level of these symptoms over time regardless of their interactions with other symptoms, we use an exponential moving average (EMA) mean function~\cite{benton2022volatility}:
    \begin{equation}
        \text{EMA}(y_{i+1,j}) = \alpha \left[y_{ij} + (1 - \alpha)y_{i-1,j} + (1 - \alpha)^2 y_{i-2,j} + \ldots + (1 - \alpha)^{k-1} y_{i-(k-1),j}\right] \enspace,
    \end{equation}    
where $\alpha = 2 / (k + 1)$ influences the smoothness of the mean. We set $k=10$. 

\subsection{Hypothesis test for dynamic covariance} \label{section:hypothesis_tests}
We determine the type of covariance between each pair of covariances using the following statistical test. With a Bayesian approach to modelling covariance processes, we obtain an estimate of the posterior distribution over the covariance process. The advantage of this is that, once we have inferred this distribution, we can perform a hypothesis test on the covariance process, allowing us to learn what type of covariance is present between a pair of variables. That is, two variables can either be i) uncorrelated, when $0 \in p(\cov_{ij}(x)\mid \textbf{x}, \textbf{Y}), \forall x$; ii) statically correlated (denoted by 'S'), if $\exists c\neq 0 \in \mathbb{R}$ such that $c \in p(\cov_{ij}(x)\mid \textbf{x}, \textbf{Y}), \forall x$; or iii) dynamically correlated (denoted by 'D'), if $ \exists c \not\in \mathbb{R}$ such that $c \in p(\cov_{ij}(x)\mid \textbf{x}, \textbf{Y}), \forall x$.

To determine whether a constant $c$ falls in the distribution $p(\cov_{ij}(x)\mid \textbf{x}, \textbf{Y})$, we determine the 95\% highest density interval of this distribution and use that instead, since no curve will have a strictly zero posterior probability. Furthermore, covariance estimates that deviate less than 0.005 from zero (or $c$) are regarded as zero (or $c$) by using the region of practical equivalence (ROPE) principle~\cite{kruschke2018rejecting}. This hypothesis test demonstrates an important advantage of estimating a distribution over the covariance process, instead of only estimating its mean. Namely, without such a distribution over the covariance process, we would not be able to use this approach to test for dynamics in the covariance. This is an important benefit of the probabilistic Wishart process compared to several common approaches, such as non-Bayesian implementations of the sliding window method and the MGARCH model.

\FloatBarrier
\subsection{Modelling of dynamic correlations between mental states}
We demonstrate how the Wishart process can be used in two different examples in studying MDD, as will be described below. In both experiments, we again sampled or optimized until convergence of all model parameters, as measured by the PSRF (see Section~\ref{section:implementation_details}). 

\subsubsection{Dynamics between mental states over time} In our first experiment, we use the Wishart process to estimate the covariance between each pair of symptoms over the course of the experiment, that is, we use the day number since the onset of the experiment as our input variable. Since we also have weekly SCL-90-R scores available, this allows us to explore whether these covariances change when the subject relapses in a depressive episode, similar to the study by Wichers et al.~\cite{wichers2016critical}. We compare our estimates to those made by a DCC-GARCH model (implemented using the R package \texttt{rmgarch}~\cite{Ghalanos_2014}. More details on its implementation are provided in Appendix~\ref{appendix:mgarch}). 

To compare the performances of the Wishart process using MCMC, variational inference and SMC, and the DCC-GARCH model, we use the following 10-fold cross validation scheme. We split the dataset evenly into 10 subsets of training and testing observations. In the first subset, we train each model on the first 36 observations and then predict the next 10 data points. In the next subset, the first 72 observations are used for training and we again predict the next 10 data points. This pattern is continued for 10 folds. We report the test log-likelihood averaged over all folds. 

\begin{figure}[!h]
    \centering
    \includegraphics[width=\textwidth]{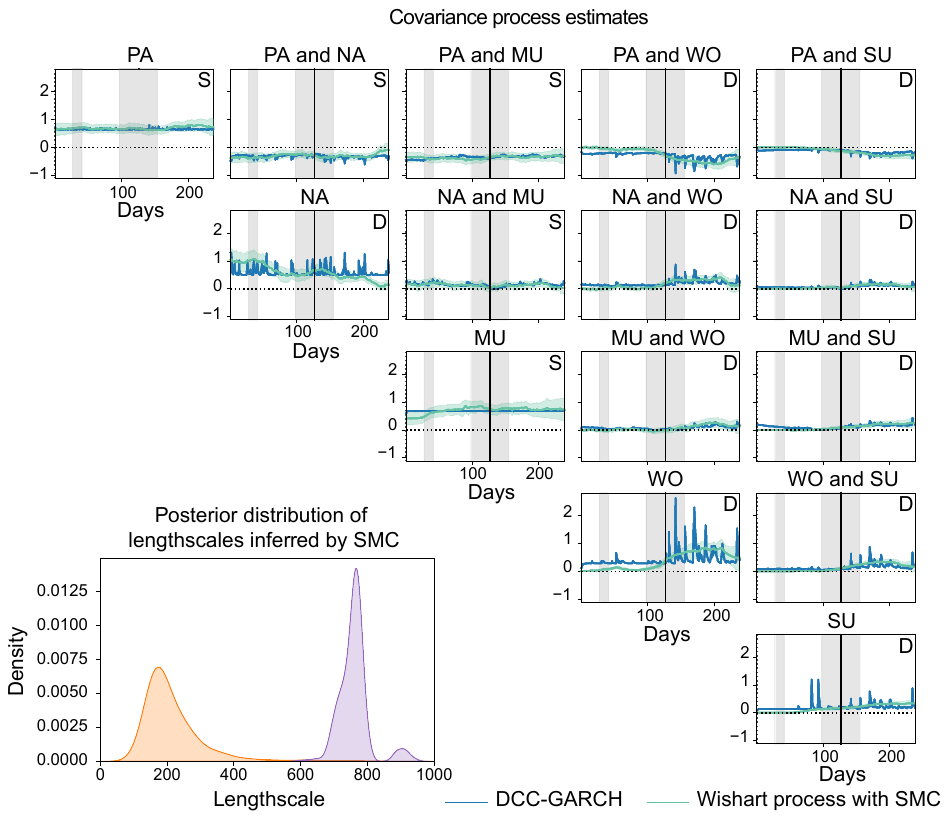}
    \caption{Estimates of the RBF and Mat\'ern 1/2 lengthscale parameters and the covariances between the five different mental states (PA = positive affect, NA = negative affect, MU = mental unrest, WO = worrying and SU = suspicious) as a function of the day number. The vertical black lines indicate the day on which the subject relapsed into depression, and the different background shades indicate different phases of the anti-depressant dose reduction scheme. We show the estimates and out-of-sample predictions of the final fold. For the Wishart process estimates, we test for dynamic covariance (D) or static covariance (S).}
    \label{fig:empirical_covariance_and_loadings}
\end{figure}

\setlength\tabcolsep{0pt}
\begin{table}[b]
    \footnotesize
    \centering
    \caption{For both the DCC-GARCH model and the Wishart process using MCMC, variational inference, or SMC, we present the mean and standard deviation of the fit to the test observations by means of the log likelihood.}
    \begin{tabular*}{0.7\linewidth}{@{\extracolsep{\fill}} lllll } \toprule
          & \multicolumn{1}{c}{DCC-GARCH} & \multicolumn{3}{c}{Wishart process} \\ \cmidrule(lr){3-5} & & \multicolumn{1}{c}{MCMC} & \multicolumn{1}{c}{VI} & \multicolumn{1}{c}{SMC} \\ \midrule
        $\text{LL}_{\text{test}}$ & -6.19 $\pm$ 2.75 & -5.29 $\pm$ 2.85 & -7.29 $\pm$ 4.21 & -5.82 $\pm$ 3.39 \\ \bottomrule
    \end{tabular*}
    \label{tab:empirical_results}
\end{table}

When modelling the covariance process as a function of time, convergence of all parameters took 5\,000\,000 MCMC burn-in steps, with a thinning of 1\,000, 61 SMC cycles with 5\,000 mutation steps per cycle, and 21\,710 VI optimization iterations. The estimates of the final fold of the Wishart process, using SMC sampling, and the DCC-GARCH model are shown in  Fig.~\ref{fig:empirical_covariance_and_loadings}. Based on visual inspection, the covariance estimates of the methods are qualitatively in agreement. Both the Wishart process and the DCC-GARCH model estimate positive affect to be negatively correlated with all other variables, and negative affect to be positively correlated with all variables except for positive affect. Moreover, previous studies~\cite{cabrieto2019objective, wichers2016critical} have found that the covariance between different mental states tends to get stronger with the increase of depressive score. As we saw in Fig.~\ref{fig:empirical_mdd_data}A, the subject relapsed into a depressive episode over time, which is, together with the different experimental phases of the venlafaxine dosage reduction, also indicated in Fig.~\ref{fig:empirical_covariance_and_loadings}. The estimates show that, overall, the covariances between the different symptoms increase in strength as the subject gets more depressed, which is in line with the findings by Wichers et al.~\cite{wichers2016critical}. In particular, we observe increased covariance between worrying and negative affect, mental unrest, and suspicion, and a decreased covariance between positive affect and worrying, and positive affect and suspicion. Finally, to determine if this covariance process is dynamic or static, we apply the in Section~\ref{section:hypothesis_tests} described hypothesis tests for dynamic covariance on the posterior distribution estimated by the Wishart process. These results suggest that all covariance pairs except for those between 'positive affect', 'negative affect', and 'mental unrest', are dynamic. By looking at the covariance function parameters inferred by SMC, also visualized in Fig~\ref{fig:empirical_covariance_and_loadings}, we observe that these dynamics in covariance are relatively slow, since the distribution over the lengthscales for the Mat\'ern 1/2 function are large compared to the input range.

In Table~\ref{tab:empirical_results}, we show the performance of the Wishart process, inferred with MCMC, variational inference or SMC, and the DCC-GARCH model for this experiment. As described in Section~\ref{sec:empirical_setup}, we evaluated each method on a 10-fold setup, where within each fold we computed the test log-likelihood over the next 10 test observations. The results show the average and standard deviation over these 10 folds, indicating that the Wishart process with MCMC or SMC, and the DCC-GARCH model, outperformed variational inference. Even though we again selected the variational inference result with the largest ELBO, the standard deviation of this method is relatively large, indicating that this method is not robust over different initializations.

\subsubsection{Dynamics between mental states as a function of venlafaxine dosage}
\begin{figure}[tb]
    \centering
    \includegraphics[width=0.9\textwidth]{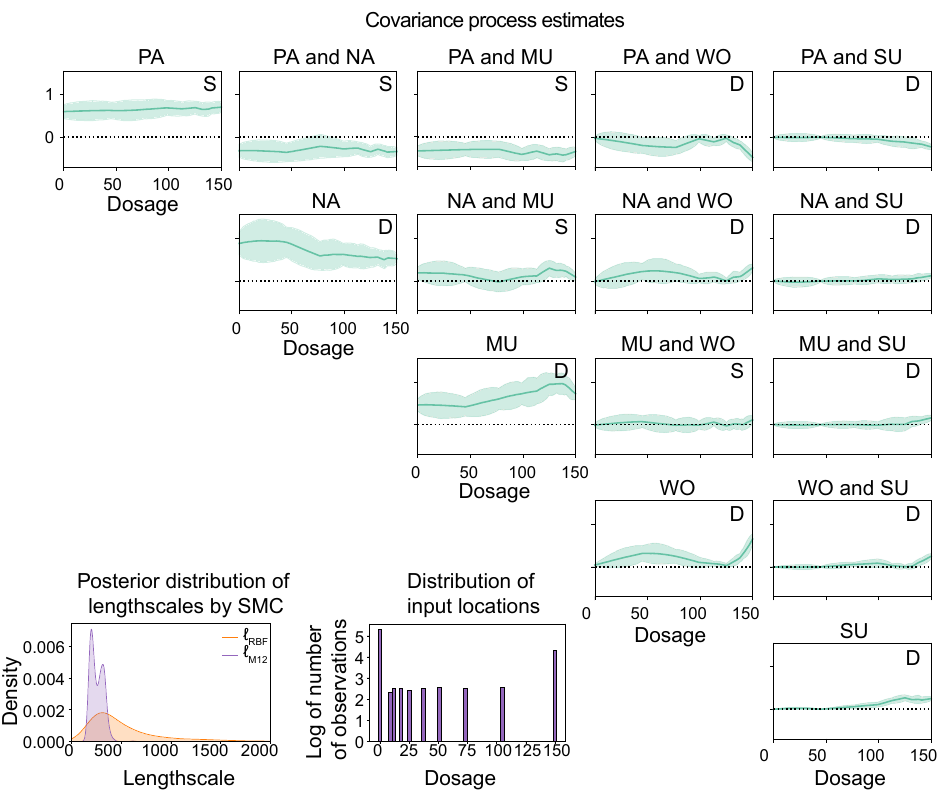}
    \caption{Estimates of the RBF and Mat\'ern 1/2 lengthscale parameters and the covariances between the five different mental states (PA = positive affect; NA = negative affect; MU = mental unrest; WO = worrying and SU = suspicious) as a function of antidepressant dosage. The bar plot indicates the amount of observations available at each input location, and we tested for dynamic (D) and static (S) covariance.}
    \label{fig:empirical_covariance_dosage}
\end{figure}
In our second application, we use venlafaxine dosage as an input variable to model the covariance between each pair of symptoms. Unlike time, venlafaxine dosage is an unevenly-spaced variable. Therefore, we can only use the Wishart process in this scenario, as the DCC-GARCH model is unable to handle unevenly-spaced input data. 

When modelling the covariance process as a function of venlafaxine dosage, convergence of all model parameters required 84 SMC cycles with 5\,000 mutation steps per cycle. The estimates by SMC are shown in Fig~\ref{fig:empirical_covariance_dosage}. Similarly to the previous demonstration, only every fourth observation is kept, resulting again in $\Observations = 369$ observations. The Wishart process can handle unevenly spaced input data, therefore this model allows us to directly model the covariance process based on the dosage reduction scheme, instead of indirectly over time. This analysis is not possible for the DCC-GARCH model, because this model requires evenly spaced input data. The resulting estimates by the Wishart process with SMC are shown in Fig.~\ref{fig:empirical_covariance_dosage}, where the input locations at which observations were measured are indicated by the red dots. As expected, these estimates are in agreement with those of Fig.~\ref{fig:empirical_covariance_and_loadings}, since the dosage was reduced over the course of the experiment. When the dosage is low, the covariances between the different mental states are generally stronger than when the dosage has not been reduced yet.

\subsubsection{Differences between dynamics in mental state correlations over time and dosage}
\begin{figure}[!h]
    \centering
    \includegraphics[width=\textwidth]{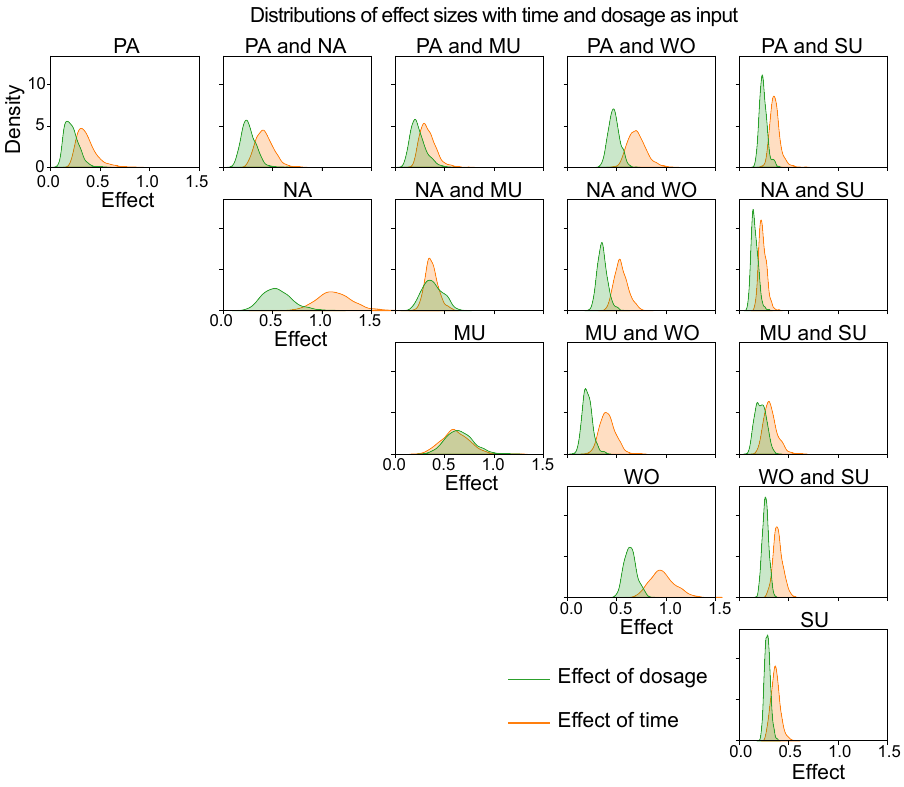}
    \caption{For each combination of mental states (PA = positive affect; NA = negative affect; MU = mental unrest; WO = worrying and SU = suspicious), and using either time or dosage as predictor variable, the distribution over the difference between the minimum and maximum of the covariance process is shown.}
    \label{fig:effect_size}
\end{figure}
Finally, we study whether there are any difference in the amount of dynamics estimated as a function of either time or dosage. We evaluate this by, for every pair of symptoms and every posterior sample, computing the distance between the minimum and maximum estimated covariance. The resulting distributions of these distances are visualized using kernel density estimation in Fig.~\ref{fig:effect_size}. This figure shows that, overall, these covariance process intervals are larger when using time as an input variable than when estimating covariance as a function of antidepressant dosage. Hence, these results imply that the effect of time on dynamic covariance is larger than the effect of dosage on dynamic covariance. This might mean that, apart from antidepressant reduction, there are other factors affecting the dynamics between the five mental states. 

\FloatBarrier

\section{Discussion}
\label{section:discussion}
Across different research fields, there is substantial interest in modelling the joint behaviour of multiple time series dynamically instead of statically. Although the Wishart process is ideally suited for this task, inference of the model parameters is challenging. Wilson and Ghahramani~\cite{wilson2010generalised} use MCMC to infer the posterior distribution of the Wishart process. Although their study showed an improved performance of modelling dynamic covariance compared to an MGARCH model, this approach is not scalable to larger number of observations and variables. Alternatively, Heaukulani and van der Wilk~\cite{heaukulani2019scalable} inferred the posterior distribution with variational inference, which scales to larger numbers of observations and variables. However, in our experiments we found that variational inference did not accurately learn the covariance function parameters, which negatively affected out-of-sample predictions of the covariance process. Moreover, both MCMC and variational inference did not give robust estimates when using composite covariance functions with multiple parameters. A similar problem has been observed in Gaussian process regression, where covariance function hyperparameters can be difficult to identify correctly~\cite{yao2022stacking, lalchand2020approximate}. In an attempt to overcome this limitation, Svensson et al.~\cite{svensson2015marginalizing} has demonstrated that Sequential Monte Carlo (SMC) can robustly marginalize over the hyperparameters. 

The SMC algorithm approximates the posterior distribution via a large amount of model parameter sets (called particles), which are initialized from their prior, and iteratively updated, weighted and resampled based on their fit to the observations. Unlike MCMC samplers, which are likely to explore high-density areas of the posterior, the different particles of SMC can cover different modes of a distribution. Therefore, SMC is more capable of dealing with multimodal distributions than MCMC and variational inference. Additionally, unlike for MCMC which is inherently sequential, large parts of the SMC algorithm can be performed in parallel, therefore speeding up the inference procedure. Since the Wishart process is constructed from GP elements, we hypothesized that inference of the Wishart process would benefit from SMC as well. Therefore, in this work, we proposed to use SMC for inference of the Wishart process parameters.

We showed that the Wishart process combined with SMC sampling indeed offers a robust approach to modelling dynamic covariance from time series data. In two simulation studies where the true covariance process was known, we found that the SMC covariance process estimates were more robust than those inferred by MCMC and variational inference, since, unlike for SMC, the results of MCMC and variational inference varied across different runs. This became especially pronounced in Simulation study 2 (see Section~\ref{section:simulation2}), where MCMC and variational inference had difficulties in inferring the covariance function hyperparameters when using composite covariance functions. Moreover, our results demonstrated that this has important consequences when we want to use the model parameters to make out-of-sample predictions. 

In general, the Wishart process is a flexible model for estimating dynamic covariance, and our work provides a key contribution in making its estimates reliable. The Wishart process allows us to model dynamic covariance over unevenly spaced data, unlike common dynamic covariance methods such as the sliding-window approach and MGARCH models. Additionally, a Bayesian approach allows for a principled approach to testing for dynamics in covariance on a single-subject using hypothesis tests. Although it is feasible to obtain a distribution over the covariance using MGARCH as well, such as through a bootstrapping approach, this is not a straightforward approach. 

The Wishart process is applicable in various fields, such as finance, where it can be used to study the interactions between stock markets~\cite{chen2022dynamic,mollah2016equity,chiang2007dynamic,karanasos2014}, neuroscience, where the interactions between brain regions are studied~\cite{lurie2020questions, kampman2024time, meng2023dynamic, cardona2015generalized} or biological systems, such as the study of social influence within animal groups \cite{sridhar2023inferring}. In our work, we demonstrated another application in biological systems, namely in psychology, where the Wishart process was used to study the covariances between mental states. With this application, we demonstrated the unique ability of the Wishart process to study covariances over unevenly spaced input variables such as medication dosage. By comparing the change in covariance across different predictors, this allows researchers to infer which predictors are the strongest drivers of changes in covariance. This is not possible with common implementations of existing approaches such as the sliding-window method or MGARCH models and therefore opens up many new possibilities for research questions across many fields. A specific example is the question whether the interactions between brain systems are mainly shaped by the developmental age of children or by other factors such as cognitive development or environmental influences~\cite{gilmore2018imaging}.

There are several important aspects that deserve consideration. First of all, although we found that SMC was able to infer the posterior more efficiently than MCMC while using a Metropolis sampler, more efficient samplers within the Gibbs cycle might be helpful, such as the Metropolis-adjusted Langevin algorithm~\cite{xifara2014langevin} or a Hamiltonian Monte Carlo algorithm~\cite{neal2010mcmc}. Moreover, although the Wishart process is a flexible model, the covariance process estimates depend on the choice of covariance function for the GP. For example, the RBF covariance function will result in relatively smooth covariance process estimates compared to the estimates when using a Mat\'ern 1/2 function. Selecting an appropriate covariance function might be challenging when the expected dynamics in covariance are unknown. In these situations, a potential solution would be to learn the covariance function from the data, for example by using the approach by Wilson and Adams~\cite{wilson2013gaussian}. Moreover, our approach with SMC currently does not scale to large numbers of observations and variables. This is because the mutation steps within SMC are parallelised, therefore storing large matrices in memory for all particles simultaneously. To improve scalability of the Wishart process with SMC, the current implementation can be augmented in several ways, of which we will provide a few suggestions here. In order to scale to larger number of observations, we could make use of inducing points~\cite{rossi2021sparse}, which was already implemented for the Wishart process using variational inference~\cite{heaukulani2019scalable}. The challenge of this approach is the trade-off between scalability and the precision of the estimates, as the performance decreases when using fewer inducing points. Another way would be to use a factored variant of the Wishart process, where a mapping from a small number of latent variables to a larger number of observed variables is learned~\cite{rowe2002multivariate, heaukulani2019scalable}. This approach works particularly well when there is a shared underlying dynamic covariance structure among a larger set of variables. Finally, when we work with evenly spaced data, we could make use of the Toeplitz structure to improve scalability ~\cite{wilson2015kernel, cunningham2008fast}. 

Currently, the Wishart process models dynamic covariance, which contains both direct as well as indirect interactions between variables. However, in certain domains, such as neuroscience, it might be more relevant to study direct interactions only, via sparse partial correlations~\cite{wang2016efficient, smith2011network}. Zero elements in a partial correlation matrix imply that there is no direct interaction between a pair of variables. Therefore, sparse partial correlations matrices can offer researchers valuable insights. Previous work \cite{hinne2014structurally} proposed to model sparse partial correlations using a G-Wishart distribution, which is, similarly to the Wishart distribution itself, a distribution over positive-definite symmetric matrices, but it also ensures that there are zero elements in the partial correlation matrices when there are no direct interactions between two variables. Extending the Wishart process to model sparse partial correlation matrices would be an interesting direction for future work.

In conclusion, the benefits of combining the Wishart process with SMC are that we can robustly estimate and out-of-sample predict covariance processes. This becomes especially pronounced when composite covariance functions, where multiple optimal parameter combinations exist, are desired. We showed how the distribution over the covariance can be used to test for dynamics, and how dynamic covariance can be modelled as a function of an unevenly spaced input. We believe that combining Wishart processes with SMC can be used to answer research questions related to dynamic covariance in different domains.

\bibliographystyle{ieeetr} 

\appendix

\section{Covariance process estimates of second simulation study} \label{appendix:estimates_study2}
\begin{figure}[!h]
    \centering
    \includegraphics[width=0.8\textwidth]{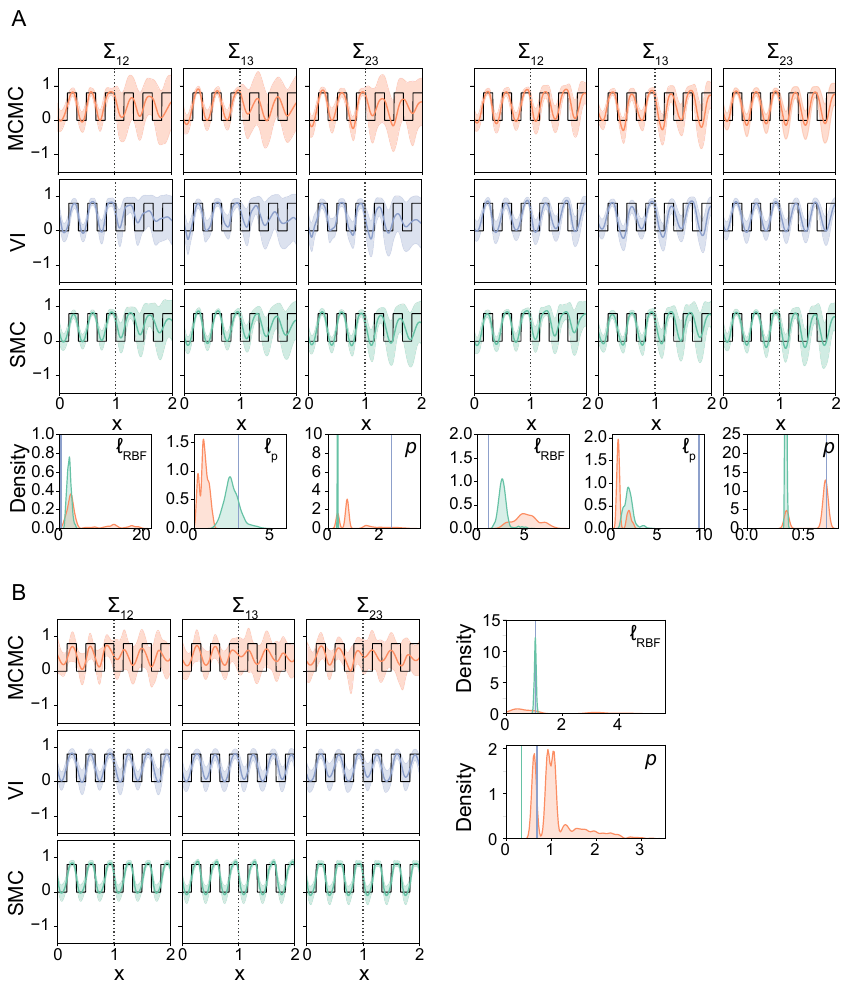}
    \caption{Covariance process estimates and out-of-sample predictions for each inference method, together with the corresponding distributions over the covariance function parameters. The true covariance process is shown in black, and out-of-sample predictions are shown after the gray dotted line. (\textbf{A}) The covariance process is modelled using a Periodic covariance function, which has two parameters: the period ($p$) and the lengthscale within each period ($\ell_{\text{RBF}}$). (\textbf{B}) Based on the same observations as in A, the covariance process is modelled using a Locally Periodic covariance function (see Eq.~\eqref{eq:lp}), which has three parameters: the period ($p$), the lengthscale within each period ($\ell_{\text{RBF}}$) and the lengthscale between periods ($\ell_{\text{p}}$).}
    \label{fig:appendix_simulation_study2_estimates}
\end{figure}
For the second simulation study, we provide the covariance process and covariance function parameter estimates based on only one dataset and covariance function. To illustrate the variety of estimates by all three inference methods, we show three more covariance process estimates in Fig.~\ref{fig:appendix_simulation_study2_estimates}, based on different sets of observations, using either a Periodic or Locally Periodic covariance function.
\section{Convergence of the covariance process samples} \label{appendix:convergence}
\begin{figure}[tb]
    \centering
    \includegraphics[width=\textwidth]{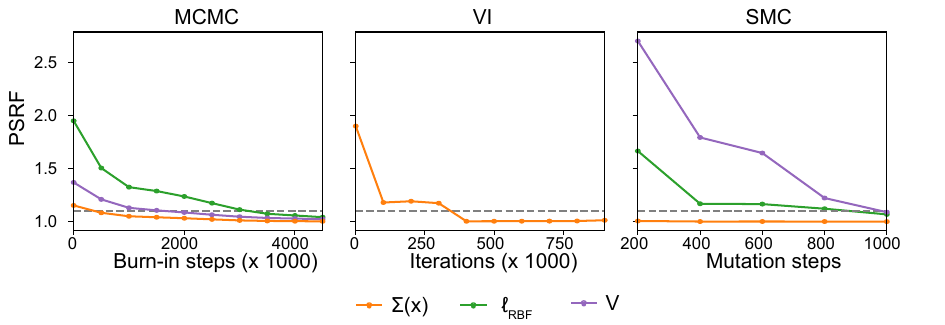}
    \caption{Convergence of the covariances, covariance function hyperparameters, and scale matrix from Section~\ref{section:simulation1}. The convergence was measured using the Potential Scale Reduction Factor (PSRF), and averaged over the elements (when applicable). Since variational inference learns point estimates of the model parameters, we only show the convergence of the covariance for variational inference. In general, convergence of the covariance requires much less burn-in steps, mutation steps, or iterations than convergence of the scale matrix and covariance function hyperparameters.}
    \label{fig:convergence}
\end{figure}
We have measured the convergence of all Wishart process parameters by means of the potential scale reduction factor between the posterior distributions resulting from different random initialisations of the model parameters. Here we found that, although the samples of the covariance process estimates converged relatively fast, convergence of the covariance function parameters and scale matrix required more burn-in steps, optimization steps, or mutation steps, as shown for the data from Simulation study 1 in Fig.~\ref{fig:convergence}.

\section{Multivariate generalised autoregressive conditional heteroscedastic models} \label{appendix:mgarch}
Multivariate generalised autoregressive conditional heteroscedastic (multivariate GARCH) models~\cite{bollerslev1988capital, bauwens2006multivariate} are an often used approach in  finance~\cite{brownlees2011practical, hansen2005forecast}. Similar to how autoregressive moving average (ARMA) models~\cite{box2015time} assume that the observations follow a Gaussian distribution and estimate current observations based on past observations and past residuals (or error terms), univariate GARCH models~\cite{bollerslev1986generalized} estimate the variance of a residual as a function of the past variances in residuals and the past residuals itself. Multivariate GARCH models estimate both the variance of a variable itself and the covariance between pairs of variables. There are several well-known versions of multivariate GARCH models. 

The Dynamic Conditional Correlation (DCC-GARCH) model by Engle~\cite{engle2002dynamic} is a variant of the multivariate GARCH model that estimates covariance from a non-linear combination of univariate GARCH models. Namely, the DCC-GARCH model specifies a univariate GARCH model $\textbf{h}_{jj} \in \mathbb{R}^\Observations (\text{for }j = 1, \ldots, \Variables)$ for every variable:
    \begin{equation}
        \textbf{H}_i = \text{diag}\left(h_{jji}^{1/2}, \ldots, h_{\Variables \Variables i}^{1/2}\right) \enspace, \qquad i=1,\ldots,\Observations \enspace.
    \end{equation}
The univariate estimates are first transformed as $u_{ij} = y_{ij} / \sqrt{h_{jji}}$ and then used to construct the symmetric positive definite matrix $\textbf{Q}_i \in \mathbb{R}^{\Variables \times \Variables}$:
    \begin{equation}
        \textbf{Q}_i = \left(1 - \alpha - \beta \right) \boldsymbol{\bar{Q}} + \alpha \textbf{u}_{i-1} \textbf{u}_{i-1} + \beta \textbf{Q}_{i-1} \enspace, \qquad i=1,\ldots,\Observations \enspace,
    \end{equation}
where $\alpha$ and $\beta$ are non-negative scalar parameters such that $\alpha + \beta < 1$, and $\boldsymbol{\bar{Q}} \in \mathbb{R}^{\Variables \times \Variables}$ is the unconditional variance matrix of $\textbf{u}_i$; it contains the variances of $\textbf{u}_i$ independent of $\textbf{u}_{i-1}$. From matrix $\textbf{Q}_j$, we now construct  $\textbf{R} \in \mathbb{R}^{\Observations \times \Variables \times \Variables}$:
    \begin{equation}
        \textbf{R}_i = \text{diag}\left(q_{jji}^{1/2}, \ldots, q_{\Variables \Variables i}^{1/2}\right) \textbf{Q}_i \text{diag}\left(q_{jji}^{1/2}, \ldots, q_{\Variables \Variables i}^{1/2}\right) \enspace, \qquad i=1,\ldots,\Observations \enspace.
    \end{equation}
Finally, the covariance estimates are constructed by combining $\textbf{H}$ and $\textbf{R}$:
    \begin{equation}
        \cov_i = \textbf{H}_i \textbf{R}_i \textbf{H}_i \enspace.
    \end{equation}
Similarly to the other multivariate GARCH variants, the DCC-GARCH model combines the outer product of the previous observations (via $\textbf{u}_j$) and the previous covariances (via $\textbf{Q}_{j-1}$) to estimate the covariance at input location $j$. The DCC-GARCH variant has $\left(\Variables+1\right) \times \left(\Variables+4\right)/2$ parameters, which is much less than the $\left( p + q \right) \left( \Variables \left( \Variables + 1 \right) / 2 \right)^2 + \Variables \left( \Variables + 1 \right) / 2 $ parameters of the original multivariate GARCH model. This makes the DCC-GARCH variant less likely to overfit.

\end{document}